\documentclass[12pt]{article}
\usepackage{amsmath,amsfonts,amssymb}
\usepackage{graphicx, hyperref}

\makeatletter
\def\numberbysection{\@addtoreset{equation}{section}
    \def\theequation{\thesection.\arabic{equation}}}
\makeatother

\numberbysection

\newcommand{\be}{\begin{equation}}
\newcommand{\ee}{\end{equation}}
\newcommand{\ba}{\begin{eqnarray}}
\newcommand{\ea}{\end{eqnarray}}
\newcommand{\nn}{\nonumber}

\begin{document}

\begin{titlepage}
\vspace{.5in}
\begin{center}

\LARGE A next-to-leading L\"uscher formula\\
\vspace{1in}
\large Diego Bombardelli \\
{\small\emph{Centro de F\'isica do Porto, Departamento de F\'isica e Astronomia, Faculdade de Ci\^encias da Universidade do Porto, Rua do Campo Alegre 687, 4169-007 Porto, Portugal; diegobombardelli@gmail.com}}

\end{center}

\vspace{.5in}

\begin{abstract}
We propose a next-to-leading L\"uscher-like formula for the finite-size corrections of the excited states energies in integrable theories. We conjecture the expressions of the corrections for both the energy and the particles' rapidities by interpreting the excited states as momenta-dependent defects. We check the resulting formulas in some simple relativistic model and conjecture those for the $AdS_{5}/CFT_{4}$ case.  
\end{abstract}

\end{titlepage}

\setcounter{footnote}{0}

\section{Introduction}\label{sec:intro}

Recently, there have been great advances in determining the $AdS/CFT$ spectrum by using integrability techniques \cite{Beisert:2010jr}. In particular, the study of the finite-volume corrections \cite{Ambjorn:2005wa, Heller:2008, Bajnok:2008bm} for the anomalous dimensions/string energies spectrum of $AdS_{5}/CFT_{4}$, has culminated in the formulation of the so-called Thermodynamic Bethe Ansatz (TBA) equations and Y-system \cite{Arutyunov:2007tc,Gromov:2009tv,Gromov:2009bc,Cavaglia:2010nm}, which in principle govern the spectrum exactly at any order of the coupling constant and the volume parameter. Very recently, the TBA equations have been reduced first to few non-linear integral (so-called FiNLIE) equations \cite{Gromov:2011cx} (see \cite{Cavaglia:2010nm,Hegedus:2009ky} for some previous developments in that direction), then to an impressively simple set of Riemann-Hilbert equations \cite{Gromov:2013pga}.

However, the highest order correction known analytically by now at weak coupling, derived by using the FiNLIE, is the 8-loop term of the Konishi operator anomalous dimension \footnote{Actually, a 9-loop result, obtained by using the methods of \cite{Gromov:2013pga}, has been presented in the talks \cite{talks}.} \cite{Leurent:2013mr}. It seems to be impossible to check analytically, at the present moment, this result by using other methods, also because, the formula provided by the generalization of the L\"uscher method \cite{Luscher, Ambjorn:2005wa, Heller:2008, Bajnok:2008bm} already reached its limit of applicability with the calculation of the 7-loop correction to the Konishi spectrum \cite{Bajnok:2012bz}.

The main aim of this paper is to start the investigation to fill this gap and to give a formula, based on the S-matrix of the theory, to calculate next-to-leading (NLO) finite-size corrections in any integrable theory.

This will be done by using the experience with the calculation of double-wrapping corrections in the vacuum of the deformed $O(4)$ $\sigma$-model and $AdS/CFT$ \cite{Ahn:2011xq}, and by considering the physical excited states as momenta-dependent defects, which twist the boundary conditions of the vacuum. We shall observe, at least in a diagonal case, that the insertion of such defects modifies the ground-state TBA equations in the same way as resulting from the standard analytical continuation of the ground-state TBA \cite{Dorey:1996re, Gromov:2009bc}.
More importantly for our purposes, the twist matrices involved in the NLO L\"uscher-like formula for the twisted vacuum \cite{Ahn:2011xq} will be replaced by the S-matrices describing the scattering between physical and virtual particles. This will give as a result the first proposal of NLO L\"uscher-like formulas for the excited states' energies and rapidities, even in a non-diagonal integrable theory.

In this paper, in particular, these formulas will be checked in some simple relativistic model - the Gross-Neveu model, the $O(4)$ $\sigma$-model and the sine-Gordon model - against the large volume expansion of their NLIEs, while the generalization for the non-relativistic case of $AdS_{5}/CFT_{4}$ will be just conjectured. We hope to come back to the actual calculation of the Konishi double-wrapping correction in the near future.

\section{Main idea}

\begin{figure}
\begin{centering}
\includegraphics[width=8cm]{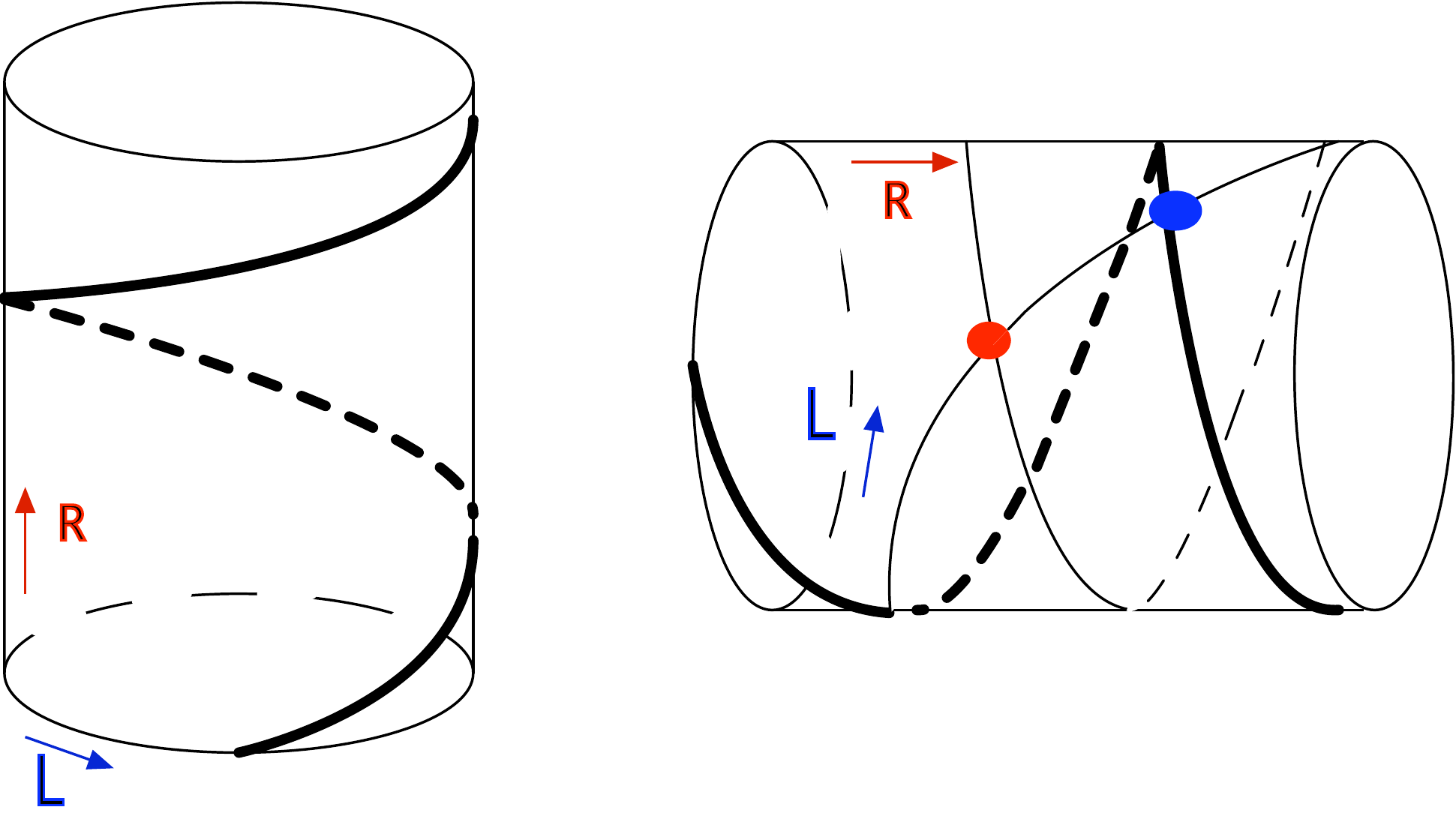}
\par\end{centering}
\caption{On the left the momentum-dependent defect is 
circling the compactified space. On the right it is located
in (Euclidean) time, and it acts as an operator on the periodic Hilbert
space of the rotated channel. The red point is a scattering between virtual particles, the blue one between a virtual particle and the defect.}
\label{Flo:dflinevsop}
\end{figure}

If we consider the physical particles as defects introduced in the compactified space of a cylinder, then we have that the usual defect transmission phase \cite{Bajnok:2004jd} is given by the S-matrix, which describes the scattering of a probe particle against $N$ excitations, that is, in a relativistic case
\be
T_{-\theta_i}(\theta)=\prod_{i=1}^NS(\theta-\theta_i)\,.
\ee
Of course, the transmission matrix in this case will depend on the rapidities of the excitations.
In this way, we are also ensured that, if the particle moves in the opposite direction, then it picks up the inverse phase:
\be
e^{ipL}=\prod_{i=1}^NS(\theta-\theta_i) \Rightarrow e^{-ipL}=\prod_{i=1}^NS^{-1}(\theta-\theta_i)=\prod_{i=1}^NS(\theta_i-\theta)\,,
\ee
where $p=\sinh\pi\theta$.
Furthermore, we have the usual relation between left and right transmission matrices
\be
T_{+\theta_i}(-\theta)=\prod_{i=1}^NS(\theta_i-\theta)=\prod_{i=1}^NS^{-1}(\theta-\theta_i)=T_{-\theta_i}^{-1}(\theta)\,,
\ee
and our transmission matrix satisfies also the defect crossing symmetry \cite{Bajnok:2004jd}:
\be
T_{-\theta_i}(\theta)=T_{+\theta_i}(i-\theta) \Leftrightarrow S(\theta-\theta_i)=S(i+\theta_i-\theta)\,.
\ee
Upon a double Wick rotation, the defect line, which defines, in our case, the asymptotic Bethe equations for the physical theory \footnote{At this point a possible reasonable objection could be: the S-matrix describes well the interaction among particles only in the limit of large $L$. Then, for finite $L$, the physical excitations should correspond to an operator describing the exact Bethe equations, with the form of the r.h.s. of (\ref{DdV}). We found only {\it a posteriori} justifications to this fact: the exact Bethe equations would reduce to the asymptotic ones if  the probe particle belongs to the rotated theory, or, in other words, particles of physical and rotated theories seem to interact only through the S-matrix, as it is confirmed by the results of this paper.}, becomes a defect operator \cite{Bajnok:2004jd, Ahn:2011xq} (see Figure \ref{Flo:dflinevsop}), which modifies the expression of the mirror \footnote{Even though in the relativistic case the space-time rotated theory is the same as the physical one, in view of the generalization to the $AdS/CFT$ case, we already start to use here the terminology related to the non-relativistic case, were the rotated theory is different from the physical one and is called mirror theory, first introduced in the first reference of \cite{Arutyunov:2007tc}.} partition function:
\be
\tilde Z(L,R)=\mbox{Tr}(e^{-\tilde H(R)L}D)\,,
\ee
where, introducing the mirror theory rapidity $\tilde\theta=\theta+i/2$, the defect operator, for a diagonal theory with single species particles, is given by \footnote{One should be careful with the analytical continuation to the mirror theory, since shifting back the integration contour to the real line, poles of the S-matrix, corresponding to bound-states between mirror and physical particles, could be met, giving additional contributions to the energy, called $\mu$-terms. We shall not consider this case through the paper.}
\be
D=\exp\left[\int\frac{d\tilde\theta}{2}\prod_{i=1}^NS(\theta_i-\tilde\theta+i/2)A(\tilde\theta)A^{\dagger}(\tilde\theta)\right]\,,
\label{D}
\ee 
with $A,A^{\dagger}$ being the Zamolodchikov-Faddeev annihilation and creation operators, respectively, in the mirror theory.
In particular, the $n$-particle matrix element of the mirror partition function can be calculated as
\be
\tilde Z(L,R)=\sum_{|\alpha_1,...,\alpha_n\rangle\in \mathcal{H}}\frac{\langle \alpha_1,...,\alpha_n|D|\alpha_1,...,\alpha_n\rangle}{\langle \alpha_1,...,\alpha_n|\alpha_1,...,\alpha_n\rangle}e^{-\tilde E_n(R)L}\,,
\ee
where
\be
D|\alpha_1,...,\alpha_n\rangle=\prod_{i=1}^NS(\theta_i-\tilde\theta_1+i/2)...S(\theta_i-\tilde\theta_n+i/2)|\alpha_1,...,\alpha_n\rangle+\mbox{permutations}\,.
\ee 
Thus, for the large volume expansion of the mirror partition function, also called cluster expansion, we have
\be
\lim_{R\rightarrow\infty}\mbox{Tr}(e^{-\tilde{H}(R)L}D)=1+\sum_{k,\alpha}\frac{\langle \alpha|D|\alpha\rangle}{\langle \alpha|\alpha\rangle} e^{
-\tilde{E}(\tilde{\theta}_{k})L}+\sum_{k\neq l,(\alpha,\beta)}\frac{\langle \alpha,\beta|D|\alpha,\beta\rangle}{\langle \alpha,\beta|\alpha,\beta\rangle}
e^{-(\tilde{E}(\tilde{\theta}_{k})+\tilde{E}(\tilde{\theta}_{l}))L}
+\dots \,,
\label{eq:ZlargeL}
\ee
where $\tilde E(\theta)=\cosh\pi\theta$. On the other hand, the physical partition function should be dominated at $R\rightarrow\infty$ by the ground state energy of the twisted physical theory: $Z(L,R)\sim e^{-R E_{0}^{d}(L)}$. However, as we shall show in what follows, expression (\ref{eq:ZlargeL}) will give the finite-size corrections to excited states' energies, hence we can say that we are calculating the ground state energy of a theory whose vacuum corresponds actually to some excited state of the original theory. In other words, the insertion into the mirror partition function of the defect operator (\ref{D}), given by the S-matrix related to a particular excited state, selects the contribution of the particular excited state energy to the large $R$ limit of the physical partition function \footnote{We thank Ryo Suzuki for a comment on this point.}.

\begin{figure}
\begin{centering}
\includegraphics[width=6cm]{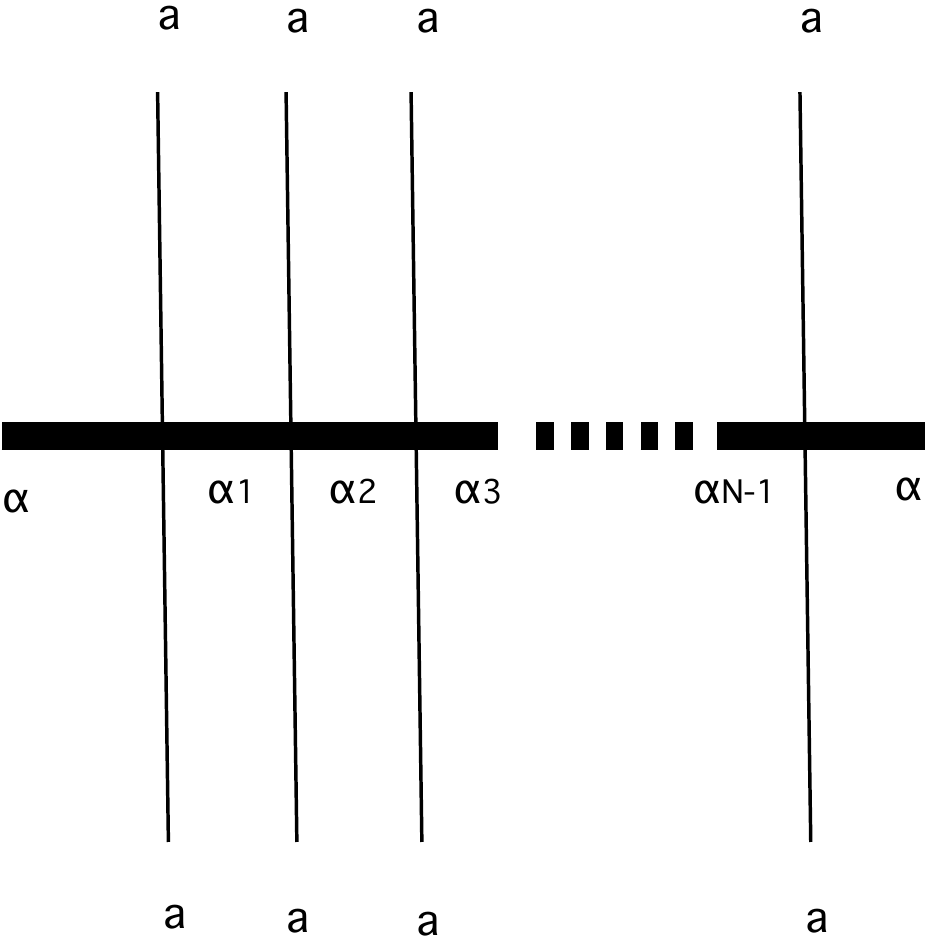}
~~~~~~~~~~~~~~~~~~~~~~~~\includegraphics[width=6cm]{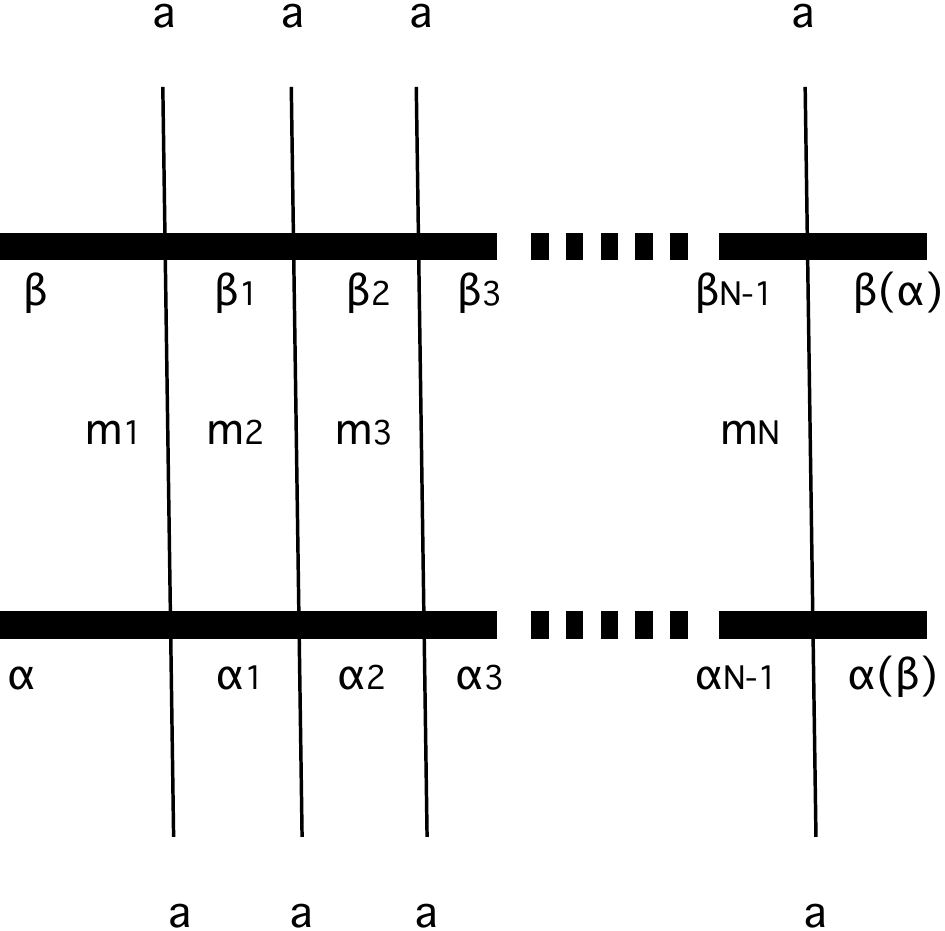}
\par\end{centering}
\caption{One-particle (left) and two-particle (right) eigenvalues of the defect operator. The horizontal thick lines represent mirror particles, whose flavors are denoted by greek indexes. The vertical thin lines with latin indexes are physical particles.}
\label{Flo:D12}
\end{figure}

In particular, in a generic non-diagonal integrable relativistic theory with particles' flavors labeled by an index $a$, the r.h.s. of (\ref{eq:ZlargeL}) can be calculated as follows (see Figure \ref{Flo:D12})
\ba
&&\hspace{-1.7cm}\frac{\langle \alpha|D_{a}|\alpha\rangle}{\langle \alpha|\alpha\rangle}=\sum_{\alpha_{1},...,\alpha_{N-1}}S_{\alpha a}^{\alpha_{1} a}(\tilde\theta-\theta_{1}+i/2)S_{\alpha_{1} a}^{\alpha_{2} a}(\tilde\theta-\theta_{2}+i/2)...S_{\alpha_{N-1} a}^{\alpha a}(\tilde\theta-\theta_{N}+i/2)\,,\label{alphaDalpha}\\
&&\hspace{-1.7cm}\frac{\langle \alpha,\beta|D_{a}|\alpha,\beta\rangle}{\langle \alpha,\beta|\alpha,\beta\rangle}= \hspace{-0.5cm}\sum_{\alpha_{1},\beta_{1},...\alpha_{N},\beta_{N}}\hspace{-0.5cm}(T_{a})_{\alpha\beta}^{\alpha_{1}\beta_{1}}(\tilde\theta_{1},\tilde\theta_{2},\theta_{1})(T_{a})_{\alpha_{1}\beta_{1}}^{\alpha_{2}\beta_{2}}(\tilde\theta_{1},\tilde\theta_{2},\theta_{2})...(T_{a})_{\alpha_{N-1}\beta_{N-1}}^{\alpha\beta}(\tilde\theta_{1},\tilde\theta_{2},\theta_{N})\nn\\
&&\hspace{-1.7cm}+\sum_{\alpha_{1},\beta_{1},...\alpha_{N},\beta_{N}}\hspace{-0.5cm}(T_{a})_{\alpha\beta}^{\alpha_{1}\beta_{1}}(\tilde\theta_{1},\tilde\theta_{2},\theta_{1})(T_{a})_{\alpha_{1}\beta_{1}}^{\alpha_{2}\beta_{2}}(\tilde\theta_{1},\tilde\theta_{2},\theta_{2})...(T_{a})_{\alpha_{N-1}\beta_{N-1}}^{\beta\alpha}(\tilde\theta_{1},\tilde\theta_{2},\theta_{N})\,,
\label{alphabetaDalphabeta}
\ea
where
\be
(T_a)_{ij}^{kl}(\tilde \theta_1,\tilde \theta_2,\theta_i)= \sum_m S_{ia}^{km}(\tilde \theta_1-\theta_i+i/2)S_{jm}^{la}(\tilde \theta_2-\theta_i+i/2)
\label{Tijkl}
\ee
and, for later purposes, we used the crossing symmetry of the defect operator.
In general, the expectation value of the non-diagonal defect operator on two generic mirror Bethe states is
\ba
&&\frac{\langle \alpha_{1},\alpha_{2},...,\alpha_{n}|D|\alpha_{1},\alpha_{2},...,\alpha_{n}\rangle}{\langle \alpha_{1},\alpha_{2},...,\alpha_{n}|\alpha_{1},\alpha_{2},...,\alpha_{n}\rangle}=\sum_{\sigma}\sum_{\underline\beta_{1},...,\underline\beta_{N-1}}(T_{a})_{\underline\alpha}^{\underline\beta_{1}}(\tilde\theta_{1},...,\tilde\theta_{n},\theta_{1})(T_{a})_{\underline\beta_{1}}^{\underline\beta_{2}}(\tilde\theta_{1},...,\tilde\theta_{n},\theta_{2})...\nn\\
&&~~~~~~~~~~~~~~~~~~~~~~~~~~~~~~~~~~~~~~~~~~~~~~~~...(T_{a})_{\underline\beta_{N-1}}^{\underline\sigma(\alpha)}(\tilde\theta_{1},...,\tilde\theta_{n},\theta_{N})\,,
\ea
where $\underline\alpha=\{\alpha_{1},\alpha_{2},...,\alpha_{n}\}$, $\underline\beta_{i}=\{\beta_{i_{1}},\beta_{i_{2}},...,\beta_{i_{n}}\}$, $\underline\sigma(\alpha)$ is any permutation of $\underline\alpha$ and
\be
(T_{a})_{\underline\alpha}^{\underline\beta}(\tilde\theta_{1},...,\tilde\theta_{n},\theta_{i})=\sum_{a_{1},...,a_{n-1}}\hspace{-0.3cm}S_{\alpha_{1}a}^{\beta_{1}a_{1}}(\tilde\theta_{1}-\theta_{i}+i/2)S_{\alpha_{2}a_{1}}^{\beta_{2}a_{2}}(\tilde\theta_{2}-\theta_{i}+i/2)...S_{\alpha_{n}a_{n-1}}^{\beta_{n}a}(\tilde\theta_{n}-\theta_{i}+i/2)\,.
\ee

On the other hand, assuming that the rapidity-dependent defect does not change the mirror Bethe equations, the TBA equations of a diagonal theory result to be modified only by the introduction of a rapidity-dependent chemical potential
\be
\mu_{\theta_i}[\rho]=R\sum_{i}\int d\theta\,\rho(\theta)\, \log[S(\theta_i-\tilde\theta+i/2)] \,,
\ee
which enters in the definition of the mirror partition function as follows
\be
\tilde Z(L,R)=\mbox{Tr}(e^{-\tilde{H}(R)L}D)=\int
d[\rho,\bar{\rho}]e^{S[\rho,\bar{\rho}]+\mu_{\theta_i}[\rho]-L\tilde{E}[\rho]} \,.
\label{eq:partitionfunction}
\ee
Following the usual procedure for the derivation of the TBA equations, the mirror free energy $f(L)=S[\rho,\bar{\rho}]+\mu_{\theta_i}[\rho]-L\tilde{E}[\rho]$ is minimized with the constraint of the mirror density equation
\be
\rho+\bar{\rho}-\frac{1}{2}\partial_{\tilde\theta}\tilde{p}=
\int d\tilde\theta'K(\tilde\theta,\tilde\theta')\rho(\tilde\theta')\,,
\ee
and the pseudo-energy $\varepsilon=\ln \bar\rho/\rho$ turns out to satisfy the following TBA equation
\be
\varepsilon(\tilde\theta)+\sum_i\log[S(\theta_i-\tilde\theta+i/2)]=\tilde{E}(\tilde\theta)L-\log(1+e^{-\varepsilon})\star K(\tilde\theta)
\,,
\label{eq:TBA}
\ee
that coincides with the standard excited states TBA equation. 

Even in a case where the physical particle was a two-particle bound-state, for instance, we would have the product $S(\theta_1-\theta+i/2)S(\theta_2-\theta+i/2)$ in the l.h.s. of (\ref{eq:TBA}), with $\theta_{1,2}$ being the rapidities of the bound-state's constituents.
For example, in the Lee-Yang model, the leading order rapidities would be $\theta_{1,2}=\pm i/3$ in our normalization, giving the same source term as in \cite{Dorey:1996re}, once crossing symmetry and unitarity are used.

Unfortunately, we did not find any argument in the particle-defects analogy in order to derive the momentum quantization condition independently from the standard analytic continuation of the TBA. Then the rapidities are not fixed by the TBA equations themselves but somehow in this approach they are put by hand. The only evidence supporting that $\theta_{i}$ are the rapidities of the physical excitations is that, as we shall see also in Section 3.3, the equation $(2n_{i}+1)i\pi=\varepsilon(\theta_i+i/2)$ at leading order at large $L$ reduces to the asymptotic Bethe equations.

So, this is just a nice check of the analogy between excited states and defects, at least in the case of a diagonal theory with single species particles. While this analysis can be easily generalized to the case with more particles' species and its extension to non-diagonal theories would be interesting, more importantly we can use this analogy to skip the solution of the excited states TBA equations in a non-diagonal case and to calculate, using equations (\ref{eq:ZlargeL}) and (\ref{alphabetaDalphabeta}), the next-to-leading L\"uscher corrections in a generic integrable theory. 

\section{L\"uscher corrections}

\subsection{Leading order}

Following the analysis of \cite{Ahn:2011xq}, we can easily write now the second term in the r.h.s. of (\ref{eq:ZlargeL}) in integral form. Then the first correction to the twisted ground-state energy, that corresponds to the energy of some particular excited state in the original model, takes the well known expression of the L\"uscher formula:  
\ba
&&E_{a}(L)=-\lim_{R\rightarrow\infty}R^{-1}\ln[\mbox{Tr}(e^{-\tilde{H}(R)L}D)]=\delta E_{a}^{(1)}(L)+O(e^{-2\tilde E L})\,,\nn\\
&&\delta E_a^{(1)}(L)=-\int \frac{d\tilde \theta}{2}\cosh\pi\tilde\theta e^{-\tilde E(\tilde \theta)L}\mbox{Tr}\prod_i^{N}S_a(\tilde \theta -\theta_i+i/2)\,,
\ea
where the trace in the integrand denotes the sum over the mirror states of (\ref{alphaDalpha})
\be
\mbox{Tr}\prod_i^{N}S_a(\tilde \theta -\theta_i+i/2)=\hspace{-0.4cm}\sum_{\alpha,\alpha_{1},...,\alpha_{N-1}}\hspace{-0.4cm}S_{\alpha a}^{\alpha_{1} a}(\tilde\theta-\theta_{1}+i/2)S_{\alpha_{1} a}^{\alpha_{2} a}(\tilde\theta-\theta_{2}+i/2)...S_{\alpha_{N-1} a}^{\alpha a}(\tilde\theta-\theta_{N}+i/2)\,,
\ee
and we recall that $a$ labels the flavor of the particular excited state we want to analyze.
In the cases we shall consider in the following sections and for the $SU(2)$ representative of the Konishi multiplet in $\mathcal{N}=4$ Super-Yang-Mills, the scattering between mirror and physical particles will be diagonal, then the expression above will simplify to \footnote{Actually, in the supersymmetric case one has to consider the super-trace:\\ $\mbox{sTr}\prod_i^{N}S_a=\sum_{\alpha}(-1)^{F_{\alpha}}\prod_{i=1}^{N}S_{\alpha a}^{\alpha a}$, with $F_{\alpha}=0,1$ for bosons/fermions respectively.}
\be
\mbox{Tr}\prod_i^{N}S_a(\tilde \theta -\theta_i+i/2)=\sum_{\alpha}\prod_{i=1}^{N}S_{\alpha a}^{\alpha a}(\tilde \theta -\theta_i+i/2)\,.
\ee  

\subsection{Next-to-leading order}

At the next-to leading order, like in \cite{Ahn:2011xq}, there is a diagonal contribution to the last term in (\ref{eq:ZlargeL}), given by
\be
-\frac{1}{2}\sum_{k,\alpha}\frac{\langle \alpha,\alpha|D|\alpha,\alpha\rangle}{\langle \alpha,\alpha|\alpha,\alpha\rangle}
e^{-2\tilde{E}(\tilde{p}_{k})L}\,,
\ee
where the mirror momenta $\tilde p_{k}$ satisfy the single-particle mirror Bethe equations $e^{i\tilde p_{k}R}=1$, and whose translation into integral form reads
\be
\delta E_{a}^{(2,1)}(L)=\int\frac{d\tilde{\theta}}{4}
\cosh\pi\tilde\theta\,e^{-2\tilde{E}(\tilde{\theta})L}\left[\mbox{Tr}\prod_i^{N}S_a(\tilde \theta-\theta_i+i/2)\right]^{2} \,.
\label{eq:E21gen}
\ee
The remaining term 
\be
\frac{1}{2}\sum_{k,l,(\alpha,\beta)}\frac{\langle \alpha,\beta|D|\alpha,\beta\rangle}{\langle \alpha,\beta|\alpha,\beta\rangle}
e^{-[\tilde{E}(\tilde{p}_{k})+\tilde{E}(\tilde{p}_{l})]L}\,,
\ee
summed over any $k,l,\alpha,\beta$, instead needs the solution of the two-particle mirror Bethe equations, that is the diagonalization of the mirror S-matrix. Taking into account the proper Jacobian in the change of variables $(k,l)\rightarrow (\tilde p_{k},\tilde p_{l})$ as in \cite{Ahn:2011xq}, we obtain
\be
\delta E_{a}^{(2,2)}(L)=-\frac{1}{2}\int d\tilde{\theta}_{1}\cosh\pi\tilde\theta_{1}\,e^{-\tilde{E}
(\tilde{\theta}_{1})L}\int \frac{d\tilde{\theta}_{2}}{2\pi}e^{-\tilde{E}
(\tilde{\theta}_{2})L}\sum_{\mu}\left[\prod_i^{N}T_a(\tilde \theta_1,\tilde \theta_2,\theta_i)\right]_{\mu}\partial_{\tilde{\theta}_{1}}\delta_{\mu}(\tilde{\theta}_{1}-\tilde{\theta}_{2}) \,,
\label{eq:E22gen}
\ee
where $\delta_{\mu}$ are the eigenvalues' phases of the S-matrix $S(\tilde \theta_1-\tilde \theta_2)$ describing the interactions of the two mirror particles, $\prod_i^{N}T_a(\tilde \theta_1,\tilde \theta_2,\theta_i)$ denotes the r.h.s. of (\ref{alphabetaDalphabeta}), the components of $T_a$ are written in (\ref{Tijkl})
and we need the expectation values of $\prod_iT_a$ on the normalized eigenvectors $|\mu\rangle$ of the mirror S-matrix:
\be
\left[\prod_i^{N}T_a(\tilde \theta_1,\tilde \theta_2,\theta_i)\right]_{\mu}\equiv\langle\mu|\prod_i^{N} T_a(\tilde \theta_1,\tilde \theta_2,\theta_i)|\mu\rangle\,.
\ee
An equivalent way to write the sum over $\mu$ in the integrand above is as follows
\be
-i\mbox{Tr}[\prod_i^{N}T_a(\tilde \theta_1,\tilde \theta_2,\theta_i) U \partial_{\tilde{\theta}_{1}}\log\Lambda(\tilde{\theta}_{1}-\tilde{\theta}_{2})U^{-1}]\,,
\ee
where $\Lambda(\tilde{\theta}_{1}-\tilde{\theta}_{2})$ is the diagonal matrix of the mirror S-matrix eigenvalues and $U$ is the change of basis matrix, that is needed here differently from the twisted vacuum case \cite{Ahn:2011xq}, since this time the S-matrix does not commute with the defect operator.

\subsection{Rapidities' corrections}

As stated above, we cannot derive the momentum quantization condition from the particle-defects analogy.
So, to understand what are the finite-size corrections for the rapidities of the physical particles, we consider the exact Bethe equations as from the standard analytic continuation of the TBA, written in terms of the pseudo-energy in a diagonal, single species case
\be
(2n_{i}+1)i\pi=\varepsilon(\theta_i+i/2)\,.
\label{EBA}
\ee
Thus, we expand at large volume, up to order $e^{-2L\tilde E}$, the r.h.s. of (\ref{EBA}):
\ba
&&\hspace{-3cm} 2i\pi n_i=iL\sinh\theta_i+\sum_{j\neq i}\log S(\theta_i-\theta_j)\nn\\
&&\hspace{-3cm}-\int d\tilde\theta\,\phi\left(\theta_i-\tilde\theta+\frac{i}{2}\right)\prod_j^{N}S\left(\theta_j-\tilde\theta+\frac{i}{2}\right)e^{-L\tilde E(\tilde\theta)}\\
&&\hspace{-3cm}+\int\frac{d\tilde\theta}{2}\phi\left(\theta_i-\tilde\theta+\frac{i}{2}\right)\prod_j^{N}S^2\left(\theta_j-\tilde\theta+\frac{i}{2}\right)e^{-2L\tilde E(\tilde\theta)}\nn\\
&&\hspace{-3cm}-\int d\tilde\theta_1d\tilde\theta_2\,\phi\left(\theta_i-\tilde\theta_1+\frac{i}{2}\right)\phi(\tilde\theta_1-\tilde\theta_2)\nn\\
&&\hspace{-0.4cm}\times\prod_{j}^{N}S\left(\theta_j-\tilde\theta_1+\frac{i}{2}\right)S\left(\theta_j-\tilde\theta_2+\frac{i}{2}\right)e^{-L(\tilde E(\tilde\theta_{1})+\tilde E(\tilde\theta_{2}))}+...\nn\,,
\label{BAEcorr}
\ea 
where $\phi(\theta)=\frac{1}{2\pi i}\frac{d}{d\theta}\ln S(\theta)$.

The first line corresponds to the asymptotic Bethe equations, while the second line is the leading order correction. In the non-diagonal case, this corresponds to the formula proposed in \cite{Bajnok:2008bm}
\be
\delta \Phi_{i,a}^{(1)}=i\int \frac{d\tilde \theta}{2\pi} e^{-\tilde E(\tilde \theta)L}\partial_{\theta_i}\mbox{Tr}\prod_j^{N}S_{a}(\theta_{j}-\tilde\theta+i/2)\,.
\ee
Now, with the help of the formulae for the energy corrections (\ref{eq:E21gen}) and (\ref{eq:E22gen}), and the three last lines of (\ref{BAEcorr}), we can guess for the non-diagonal case the following next-to-leading order rapidities' corrections:
\be
\delta \Phi_{i,a}^{(2,1)}=-\frac{i}{4}\int \frac{d\tilde{\theta}}{2\pi}
e^{-2\tilde{E}(\tilde{\theta})L}\partial_{\theta_i}\left[\mbox{Tr}\prod_j^{N}S_a(\theta_{j}-\tilde\theta+i/2)\right]^{2} \,,
\label{eq:phi21gen}
\ee
and
\be
\delta \Phi_{i,a}^{(2,2)}=\frac{i}{2}\int \frac{d\tilde{\theta}_{1}}{2\pi}e^{-\tilde{E}
(\tilde{\theta}_{1})L}\int \frac{d\tilde{\theta}_{2}}{2\pi}e^{-\tilde{E}
(\tilde{\theta}_{2})L}\partial_{\theta_i}\sum_{\mu}\left[\prod_j^{N}T_a(\theta_j,\tilde\theta_1,\tilde\theta_2)\right]_{\mu}\partial_{\tilde{\theta}_{1}}\delta_{\mu}(\tilde{\theta}_{1}-\tilde{\theta}_{2}) \,.
\label{eq:phi22gen}
\ee
Equations (\ref{eq:E21gen}), (\ref{eq:E22gen}), (\ref{eq:phi21gen}) and (\ref{eq:phi22gen}) are the main results of this paper. Now we shall check them against the large volume expansion of some relativistic models' well known NLIEs. 

\section{Comparison with NLIE}

\subsection{Gross-Neveu model}

The S-matrix of the chiral $SU(2)$ Gross-Neveu model is
\be
S(\theta)=\frac{S_{0}(\theta)}{(\theta-i)}\hat{S}(\theta)\,, \qquad
\hat{S}(\theta)=\theta\,\mathbb{I}-i\,\mathbb{P} \,,
\label{Smatrix}
\ee
where the scalar factor is
\be
S_{0}(\theta)=i\frac{\Gamma(\frac{1}{2}-\frac{i\theta}{2})\Gamma(\frac{i\theta}{
2})}
{\Gamma(\frac{1}{2}+\frac{i\theta}{2})\Gamma(-\frac{i\theta}{2})}\,.
\label{scalarfactor}
\ee
In the $U(1)$ sector, the DdV equation can be written as \cite{Gromov:2008gj}
\be
g(x)=e^{iL\sinh\pi x} \prod_{j=1}^N S_0(x-\theta_j)\exp\{2i\mbox{Im}K_0^{-}*\log[1+g^{+}]\}\,,
\label{DdV}
\ee
where $K_0(x)=\frac{1}{2\pi i}\partial_x \log S_0(x)$ and $f^{\pm}(x)=f(x\pm i/2)$ \footnote{We are using through all this Section the notations of \cite{Gromov:2008gj}.}. Actually, since in the formula for the energy
\be
E=\sum_j \cosh(\pi\theta_j)-\frac{1}{2}\int d\tilde\theta\cosh(\pi\tilde\theta) [\log(1+g^{+})+\log(1+1/\bar g^{-})]
\ee
one needs $g^{+}$ and its complex conjugate $1/\bar g^{-}$, we find convenient to define $A\equiv g^+$ and to use the following non-linear integral equation (NLIE):
\be
A(x)=e^{-L\cosh \pi x} \prod_{j=1}^N S_0(x-\theta_j+i/2)\exp[K_0*\log(1+A)-K_0^{++}*\log(1+\bar A)-\log(1+\bar A)]\,,
\ee
where one has to take the principal value of the convolution involving $K_{0}^{++}$ and the last term in the exponential is due to its ($-1/2$) residue in $y=x$. 

Now, in order to calculate the leading energy correction, we consider the leading order of $A(x)$ in the large $L$ expansion, 
\be
A_0(x)=e^{-L\cosh \pi x} \prod_{j=1}^N S_0(x-\theta_j+i/2)\,;\quad \bar A_0(x)=e^{-L\cosh \pi x} \prod_{j=1}^N S_0^{-1}(x-\theta_j-i/2)\,,
\ee
which gives the well known L\"uscher term
\ba
&&\delta E^{(1)}=-\frac{1}{2}\int d\tilde\theta\cosh\pi\tilde\theta \left[\prod_{j=1}^N S_0(\tilde\theta-\theta_j+i/2)+\prod_{j=1}^N S_0^{-1}(\tilde\theta-\theta_j-i/2)\right]e^{-L\cosh\pi\tilde\theta}\nn\\
&&=-\frac{1}{2}\int d\tilde\theta\cosh\pi\tilde\theta\,\mbox{Tr}\prod_{j=1}^N S(\tilde\theta-\theta_j+i/2)\,.
\ea
At the next-to-leading order, the function $A(x)$ is
\ba
&&\hspace{-0.8cm}A_1(x)=e^{-L\cosh \pi x} \prod_{j=1}^N S_0(x-\theta_j+i/2)[K_0*\log(1+A)-K_0^{++}*\log(1+\bar A)-\log(1+\bar A)]\,,\nn\\
&&\hspace{-0.8cm}\bar A_1(x)=e^{-L\cosh \pi x} \prod_{j=1}^N S_0^{-1}(x-\theta_j-i/2)[K_0*\log(1+\bar A)-K_0^{--}*\log(1+A)-\log(1+A)]\,,\nn
\ea
that implies the following energy correction 
\ba
&&\delta E^{(2,1)}+\delta E^{(2,2)}=\frac{1}{4}\int d\tilde\theta\cosh\pi\tilde\theta\left[ S^+(\theta)+\frac{1}{S^{-}(\theta)}\right]^2e^{-2L\cosh\pi\tilde\theta}\nn\\
&&-\frac{1}{2}\int d\tilde\theta_1 d\tilde\theta_2  \cosh\pi\tilde\theta_1e^{-L(\cosh\pi\tilde\theta_1+\cosh\pi\tilde\theta_2)}\left\{K_0(\tilde\theta_1-\tilde\theta_2)\left[  S^+(\tilde\theta_1)S^+(\tilde\theta_2)+\frac{1}{S^{-}(\tilde\theta_1)S^{-}(\tilde\theta_2)}\right]\right.\nn\\
&&\left.-K_0(\tilde\theta_1-\tilde\theta_2+i)\frac{S^+(\tilde\theta_1)}{S^-(\tilde\theta_2)}-K_0(\tilde\theta_1-\tilde\theta_2-i)\frac{S^+(\tilde\theta_2)}{S^-(\tilde\theta_1)}\right\}\,,
\label{eq:E2GN}
\ea
where $S^{\pm}(\theta)\equiv\prod_{j=1}^N S_0(\theta-\theta_j\pm i/2)$. We checked that the integrands of (\ref{eq:E2GN}) are the same of equations (\ref{eq:E21gen}) and (\ref{eq:E22gen}) with the S-matrix given by (\ref{Smatrix}) and (\ref{scalarfactor}), with flavor label $a=1$, for any values of the physical rapidities. The agreement between (\ref{eq:E22gen}) and the second integral of (\ref{eq:E2GN}) is not automatic: in order to show it, one has to use the identities
\be
S_{0}^{+}(\theta)S_{0}^{-}(\theta)=\frac{\theta-\frac{i}{2}}{\theta+\frac{i}{2}}\,,\quad K_{0}(\theta)+K_{0}^{\pm\pm}(\theta)=\frac{1}{2\pi\theta(\theta\pm i)}\,.
\label{identities}
\ee
Thus, at least for the states belonging to the $U(1)$ sector of the Gross-Neveu model we have full agreement with our formulas (\ref{eq:E21gen}), (\ref{eq:E22gen}). 

On the other hand, the exact Bethe equations are given by
\be
g(\theta_{i})=-1\,.
\label{exactE}
\ee
Then, at the zeroth order at large $L$, it gives the ABA
\be
g_0(x)=e^{iL\sinh \pi x} \prod_{j=1}^N S_0(x-\theta_j)=-1\,,\ x=\theta_i\ \Rightarrow\ 2in_k\pi=iL\sinh\pi\theta_k+\sum_{j=1}^N \log S_0(\theta_i-\theta_j)\,.
\ee
At the first order, (\ref{exactE}) gives the Bajnok-Janik formula \cite{Bajnok:2008bm}
\be
g_1(x)=g_0(x)+2i\mbox{Im}[K_0^-*\log(1+g_0^{+})]=-1\,,\ x=\theta_i\ \Leftrightarrow 2in_i\pi=BY_i+\delta\Phi_i^{(1)}\,,
\ee
where
\ba
&&\delta\Phi_i^{(1)}=2i\int d\tilde\theta \mbox{Im}[K_0^-(\theta_i-\tilde\theta)*\log(1+g_0^{+}(\tilde\theta))]=\int d\tilde\theta \left[K_0^-(\theta_i-\tilde\theta) S^+(\tilde\theta)-\frac{K_0^+(\theta_i-\tilde\theta)}{S^-(\tilde\theta)}\right]\nn\\
&&=i\int \frac{d\tilde\theta}{2\pi} \partial_{\theta_i}\mbox{Tr}\prod_j^{N} S(\theta_j-\tilde\theta+i/2)\,.
\ea
At the second order, we obtain
\be
g_2(x)=g_0(x)+2i\mbox{Im}[K_0^-*\log(1+g_1^{+})]=-1\,,\ x=\theta_i\ \Leftrightarrow 2in_i\pi=BY_i+\delta\Phi_i^{(1)}+\delta\Phi_i^{(2,1)}+\delta\Phi_i^{(2,2)}\,,
\ee
where the NLO corrections result to be
\ba
&&\hspace{-1.2cm}\delta\Phi_i^{(2,1)}
=\frac{1}{2}\int d\tilde\theta K_0^+(\theta_i-\tilde\theta) \left[\frac{1}{(S^-)^2(\tilde\theta)}+\frac{S^+(\tilde\theta)}{S^-(\tilde\theta)}\right]-K_0^-(\theta_i-\tilde\theta)\left[(S^+)^2(\tilde\theta)+\frac{S^+(\tilde\theta)}{S^-(\tilde\theta)}\right]
\label{eq:phi21GN}\\
&&\hspace{-1.2cm}\delta\Phi_i^{(2,2)}=-\int d\tilde\theta_1 d\tilde\theta_2 K_0^+(\theta_i-\tilde\theta_1) \left[\frac{K_0(\tilde\theta_1-\tilde\theta_2)}{S^-(\tilde\theta_1)S^-(\tilde\theta_2)}-K_0(\tilde\theta_1-\tilde\theta_2-i)\frac{S^+(\tilde\theta_2)}{S^-(\tilde\theta_1)}\right]\nn\\
\hspace{-1cm}&&-K_0^-(\theta_i-\tilde\theta)\left[K_0(\tilde\theta_1-\tilde\theta_2)S^+(\tilde\theta_1)S^+(\tilde\theta_2)-K_0(\tilde\theta_1-\tilde\theta_2+i)\frac{S^+(\tilde\theta_1)}{S^-(\tilde\theta_2)}\right]\,.
\label{eq:phi22GN}
\ea
While in order to show the agreement between (\ref{eq:phi21GN}) and (\ref{eq:phi21gen}), in the case of $a=1$ for instance, it is enough to use the unitarity of $S_{0}$, the matching between (\ref{eq:phi22GN}) and (\ref{eq:phi22gen}) requires the use of identities (\ref{identities}). 
Still, the integrands of (\ref{eq:phi22gen}) and (\ref{eq:phi22GN}) differ by a term that is, for example in the one-particle case, equal to
\be
-\frac{K_{0}^{++}(\tilde\theta_{1}-\tilde\theta_{2})-K_{0}^{--}(\tilde\theta_{1}-\tilde\theta_{2})}{8\pi i(\tilde\theta_{1}-\theta_{1}+i/2)(\tilde\theta_{2}-\theta_{1}+i/2)}\,,
\ee  
where $\theta_{1}$ is the rapidity single physical excitation.
This term is antisymmetric under the exchange $\tilde\theta_{1}\leftrightarrow \tilde\theta_{2}$, then we have complete agreement upon the double integration in $\tilde\theta_{1}$, $\tilde\theta_{2}$.

\subsection{Sine-Gordon and $O(4)$ $\sigma$-model}

We checked our formulas also in the cases of sine-Gordon, whose limit $\nu\rightarrow\infty$ is the Gross-Neveu model, and the $O(4)$ $\sigma$-model ($SU(2)$ principal chiral model), which is basically the tensor product of two Gross-Neveu models.

In the first case, it is enough to replace the S-matrix (\ref{Smatrix}) by
\be
S(\theta,\nu)=S_{0}(\theta,\nu)\hat{S}(\theta,\nu)\,, \qquad
\hat{S}(\theta,\nu)=\left(
\begin{array}{cccc}
1&0&0&0\\
0&\frac{\sinh\left(\frac{\pi}{\nu}\theta\right)}{\sinh\left(\frac{\pi}{\nu}(\theta-i)\right)}&\frac{-\sinh\left(\frac{\pi}{\nu}i\right)}{\sinh\left(\frac{\pi}{\nu}(\theta-i)\right)}&0\\
0&\frac{-\sinh\left(\frac{\pi}{\nu}i\right)}{\sinh\left(\frac{\pi}{\nu}(\theta-i)\right)}&\frac{\sinh\left(\frac{\pi}{\nu}\theta\right)}{\sinh\left(\frac{\pi}{\nu}(\theta-i)\right)}&0\\
0&0&0&1
\end{array}
\right)\,,
\ee
and the scalar factor by
\be
S_{0}(\theta,\nu)=-i\exp i\int_{0}^{\infty}d\omega\frac{\sin\omega x}{\omega}\frac{\sinh\left(\frac{\nu-1}{2}\omega\right)}{\cosh\left(\frac{\omega}{2}\right)\sinh\left(\frac{\nu}{2}\omega\right)}\,.
\ee
All the other quantities follow from these in both sides, L\"uscher and NLIE, of the comparison.
In the case of the $O(4)$ $\sigma$-model, instead, one needs to use
\be
S(\theta)=\frac{S_{0}^{2}(\theta)}{(\theta-i)^{2}}\hat{S}(\theta)
\otimes\hat{S}(\theta)\,, \qquad
\hat{S}(\theta)=\theta\,\mathbb{I}-i\,\mathbb{P} \,,
\ee  
where the scalar factor $S_{0}$ is the same as that for the Gross-Neveu model (\ref{scalarfactor}).
In more detail, the main formula for the double-wrapping correction (\ref{eq:E22gen}) becomes
\ba
&&\delta E_{a}^{(2,2)}(L)=i\int d\tilde{\theta}_{1}\,\cosh\tilde\theta_{1} e^{-\tilde{E}
(\tilde{\theta}_{1})L}\int \frac{d\tilde{\theta}_{2}}{2\pi}e^{-\tilde{E}
(\tilde{\theta}_{2})L}\mbox{Tr}\left[\prod_{i}^{N}T_{a}^{SU(2)}(\tilde\theta_{1},\tilde\theta_{2},\theta_{i})\right]\nn\\
&&\times\,\mbox{Tr}\left[\prod_i^{N}T_a^{SU(2)}(\tilde \theta_1,\tilde \theta_2,\theta_i)U\partial_{\tilde{\theta}_{1}}\log\Lambda_{SU(2)}(\tilde{\theta}_{1}-\tilde{\theta}_{2})U^{-1}\right] \,,
\ea
where $T_{a}^{SU(2)}$, $U$ and $\Lambda_{SU(2)}$ are the same quantities involved in the $SU(2)$ chiral Gross-Neveu model, while the NLIE for the $U(1)$ sector changes to \cite{Gromov:2008gj}
\be
A(x)=e^{-L\cosh\pi x} \prod_{j=1}^N S_0(x-\theta_j+i/2)\exp\left[K_0*\log\frac{1+A}{1-|A|}-K_0^{++}*\log\frac{1+\bar A}{1-|A|}-\log\frac{1+\bar A}{1-|A|}\right]\,.
\ee
In both cases, we obtained exact agreement between our NLO L\"uscher-like formulas and the expansions of the well known NLIEs \cite{Destri:1987ze,Gromov:2008gj}, at least for excitations belonging to the $U(1)$ sectors, in the same way as for the Gross-Neveu model.
In the sine-Gordon case we needed to use a generalized version of the identities (\ref{identities}):
\be
S_{0}^{+}(\theta,\nu)S_{0}^{-}(\theta,\nu)=\frac{\sinh\frac{\pi}{\nu}(\theta-\frac{i}{2})}{\sinh\frac{\pi}{\nu}(\theta+\frac{i}{2})}\,,\quad K_{0}^{+}(\theta,\nu)+K_{0}^{-}(\theta,\nu)=\frac{\coth\frac{\pi}{\nu}(\theta-\frac{i}{2})-\coth\frac{\pi}{\nu}(\theta+\frac{i}{2})}{2\nu i}\,.
\ee
Clearly, it would be interesting to perform the same check for other sectors, and for other integrable relativistic theories. In the case of complex solutions, bound-states etc. there can be additional contributions which are not taken into account in our analysis.

\section{Conjectures for AdS/CFT}

On the basis of the previous results, it is not so difficult to guess the following formula as the usual quadratic contribution to the next-to-leading energy corrections for excited states in $AdS_{5}/CFT_{4}$
\be
\delta E_{a}^{(2,1)}(L)=\frac{1}{2}\sum_{Q=1}^{\infty}\int\frac{d\tilde{p}}{2\pi}
e^{-2\tilde{E}_{Q}(\tilde{p})L}\left[\mbox{sTr}\prod_i^{N}S_a^{Q,1}(\tilde p,p_i)\right]^{2} \,,
\label{eq:E21ads}
\ee
where
\be
\tilde E_{Q(\tilde p)}=2\mbox{arcsinh}\left(\frac{\sqrt{Q^{2}+\tilde p^{2}}}{2g}\right)
\ee
is the mirror dispersion relation depending on the mirror momentum $\tilde p$, and sTr denotes the super-trace, that in the simple case of diagonal mirror-physical scattering, like in the $SU(2)$ sector for instance, becomes
\be
\mbox{sTr}S_a^{Q,1}=\sum_{b}(-1)^{F_{b}}\prod_{i}^{N}(S^{Q,1}(\tilde p,p_{i}))_{ba}^{ba}\,.
\ee 
$S^{Q,1}$ is the S-matrix describing the scattering between a mirror bound-state with charge $Q$ and a single physical particle \cite{Arutyunov:2008zt}, where $S$ is the $AdS_{5}/CFT_{4}$ S-matrix \cite{Beisert:2005tm}: 
\be
S=S_{0}^{2}(S_{SU(2|2)}\otimes S_{SU(2|2)})\,.
\ee
So, (\ref{eq:E21ads}) can be written in terms of the $SU(2|2)$-invariant S-matrix as 
\be
\delta E_{a}^{(2,1)}(L)=\frac{1}{2}\sum_{Q=1}^{\infty}\int\frac{d\tilde{p}}{2\pi}
e^{-2\tilde{E}_{Q}(\tilde{p})L}\left[\prod_i^{N}S_{0}^{Q,1}(\tilde p,p_i)\sum_{b}(-1)^{F_{b}}\prod_i^{N}(S_{SU(2|2)}^{Q,1})_{ba}^{ba}(\tilde p,p_i)\right]^{4} \,.
\label{eq:E21ads2}
\ee
It matches also the prediction from the large volume expansion of the excited states TBA
\be
\delta E_{TBA}=-\sum_{Q=1}^{\infty}\int \frac{d\tilde{p}}{2\pi}Y_{Q}^{(0)}(\tilde p,p_{i})-\frac{1}{2}(Y_{Q}^{(0)})^{2}(\tilde p,p_{i})+...\,,
\ee
where $Y_{Q}^{(0)}$ is the asymptotic solution \cite{Gromov:2009tv} for the $Y_{Q}$-functions of the excited states TBA equations \cite{Gromov:2009bc}.
The more complicated contribution reads
\ba
&&\delta E_{a}^{(2,2)}(L)=2i\sum_{Q_{1},Q_{2}}\int\frac{d\tilde{p}_{1}}{2\pi}\, e^{-\tilde{E}_{Q_{1}}
(\tilde{p}_{1})L}\int\frac{d\tilde{p}_{2}}{2\pi}e^{-\tilde{E}_{Q_{2}}
(\tilde{p}_{2})L}\mbox{sTr}\left[\prod_i^{N}T_a^{Q_{1},Q_{2}}(\tilde p_1,\tilde p_2,p_i)\right]\nn\\
&&\times\,\mbox{sTr}\left\{\prod_i^{N}T_a^{Q_{1},Q_{2}}(\tilde p_1,\tilde p_2,p_i)U_{SU(2|2)}^{Q_{1},Q_{2}}\left[\partial_{\tilde{p}_{1}}\log\Lambda_{SU(2|2)}^{Q_{1},Q_{2}}(\tilde{p}_{1},\tilde{p}_{2})\right](U_{SU(2|2)}^{Q_{1},Q_{2}})^{-1}\right\} \,,
\label{eq:E22ads}
\ea
where 
\be
(T_a^{Q_{1},Q_{2}})_{ij}^{kl}(\tilde p_1,\tilde p_2,p_i)=S_{0}^{Q_{1},1}(\tilde p_1,p_i)S_{0}^{Q_{2},1}(\tilde p_2,p_i)\sum_m (S_{SU(2|2)}^{Q_{1},1})_{ia}^{km}(\tilde p_1,p_i)(S_{SU(2|2)}^{Q_{2},1})_{jm}^{la}(\tilde p_2,p_i)\,.
\ee
Moreover, $\Lambda_{SU(2|2)}^{Q_{1},Q_{2}}$ is the diagonalized $SU(2|2)$-invariant S-matrix for the scattering between two generic mirror bound-states \cite{Arutyunov:2009mi}, and $U_{SU(2|2)}^{Q_{1},Q_{2}}$ is its change of basis matrix. In order to calculate the finite-size correction to the 8-loop anomalous dimension of the Konishi operator one needs just formulas (\ref{eq:E21ads2}) and (\ref{eq:E22ads}), with $a=1$, $j=1,2$, $p_{1}=-p_{2}=2\pi/3$, together with the leading order formulas of \cite{Bajnok:2008bm} for energy and rapidities' corrections, expanded to $g^{16}$.
A preliminary check of (\ref{eq:E22ads}) can be done by isolating the scalar factor from $\Lambda_{SU(2|2)}$ and choosing $a=1$; then one gets
\ba
&&\delta E_{1}^{(2,2)}(L)=-\sum_{Q_{1},Q_{2}}\int\frac{d\tilde{p}_{1} d\tilde{p}_{2}}{2\pi}\, e^{-(\tilde{E}_{Q_{1}}
(\tilde{p}_{1})+\tilde{E}_{Q_{2}}
(\tilde{p}_{2}))L}K_{sl(2)}^{Q_{1}Q_{2}}(\tilde p_{1},\tilde p_{2})\\
&&~~~~~~~~~~~~~~~~~~~~~~~~~~~~~~~~~~~~~~~~~~~~~~\times\,\mbox{sTr}\prod_i^{N}(S^{Q_{1},1}_{1})(\tilde p_1,p_i)\mbox{sTr}\prod_i^{N}(S^{Q_{2},1}_{1})(\tilde p_2,p_i)\nn\,,
\ea
that matches the result from the TBA
\be
\delta E^{(2,2)}_{TBA}=-\sum_{Q_{2}=1}^{\infty}\int\frac{d\tilde{p}_{2}}{2\pi}Y_{Q_{2}}^{(0)}(\tilde p_{2},p_{i})Y_{Q_{2}}^{(1)}(\tilde p_{1},\tilde p_{2},p_{i})\,,
\ee
where one considers the only contribution in the NLO solution $Y_{Q_{2}}^{(1)}$ given by the dressing kernel 
\be
Y_{Q_{2}}^{(1)}(\tilde p_{1},\tilde p_{2},p_{i})=Y_{Q_{1}}^{(0)}(\tilde p_{1},p_{i})\star K_{sl(2)}^{Q_{1}Q_{2}}(\tilde p_{1},\tilde p_{2})+...\,.
\ee
 
It will be also possible to go further in the number of loops - up to 11 in particular, since at 12 loops triple-wrapping effects appear - by considering the following conjectures for the corrections of the rapidities
\ba
&&\delta \Phi_{i}^{(2,1)}=\frac{i}{4}\sum_{Q}\int\frac{d\tilde{p}}{2\pi}
e^{-2\tilde{E}_{Q}(\tilde{p})L}\partial_{\tilde p}\left[\mbox{sTr}(\prod_j^{N}S_a^{Q,1}(\tilde p,p_j)\right]^{2} \,,
\label{eq:aba21ads}\\
&&\delta \Phi_{i}^{(2,2)}=-2\sum_{Q_{1},Q_{2}}\int\frac{d\tilde{p}_{1}}{2\pi}\, e^{-\tilde{E_{Q_{1}}}
(\tilde{p}_{1})L}\int\frac{d\tilde{p}_{2}}{2\pi}e^{-\tilde{E_{Q_{2}}}
(\tilde{p}_{2})L}\mbox{sTr}\left[\partial_{\tilde p_1}\prod_i^{N}T_a^{Q_{1},Q_{2}}(\tilde p_1,\tilde p_2,p_i)\right]\nn \\
&&\times\,\mbox{sTr}\left\{\prod_i^{N}T_a^{Q_{1},Q_{2}}(\tilde p_1,\tilde p_2,p_i)U_{SU(2|2)}^{Q_{1},Q_{2}}\left[\partial_{\tilde{p}_{1}}\log\Lambda_{SU(2|2)}^{Q_{1},Q_{2}}(\tilde{p}_{1},\tilde{p}_{2})\right](U_{SU(2|2)}^{Q_{1},Q_{2}})^{-1}\right\} \,.
\label{eq:aba22ads}
\ea

We do not consider here the contributions from possible $\mu$-terms, which could appear at this or higher loops.

\section{Summary and Outlook}\label{sec:discussion}

We proposed next-to-leading L\"uscher-like formulas, involving just the S-matrix and the dispersion relation of the theory, for generic integrable theories, also for the finite-size corrections to the rapidities of the physical particles/excited states.

In doing this, we were helped by a similar derivation for the twisted ground-states \cite{Ahn:2011xq}, and by the idea that the excited states could be seen as defects, at least from the point of view of the interaction with the virtual/mirror particles.

We found full agreement with the large volume expansion of the NLIE in the cases of few relativistic theories and conjectured a generalization of our formulae for the case of $AdS_{5}/CFT_{4}$.
Due to the complication of the calculations, we leave the check against the known results for the Konishi operator, the attempt to calculate its anomalous dimension up to 11 loops and the analysis of possible $\mu$-terms for future work.

A simpler check in the $AdS/CFT$ case could be a comparison between the results from our formulas at strong coupling, with a single physical particle and at most two mirror single particles ($Q_{1}=Q_{2}=1$) and a mirror bound-state with $Q=2$, against the F-term already calculated at any order for generic string solutions moving in $AdS_{3}\times S^{1}$ \cite{Gromov:2009tq} or the giant magnon \cite{Heller:2008, Abbott:2011tp}.

Other checks could be also performed in theories different form those analyzed in this paper or in other sectors of the excited states' spectrum.

Finally, it would be absolutely interesting to study possible generalizations of this approach to higher orders in the large volume expansion, in order to investigate possible simplifications or regular patterns, which could guide us to the exact solution of the spectra in particular sectors of interesting integrable theories, such as various examples of $AdS_{d+1}/CFT_{d}$. 

\section*{Acknowledgments}

I am especially indebted to Zoltan Bajnok for enlightening insights, discussions and useful comments on the manuscript. I am grateful to the Wigner Research Centre in Budapest, where part of this work was performed, for the warm hospitality. I thank also Davide Fioravanti, Rafael Nepomechie, Francesco Ravanini, Simone Piscaglia and Emilio Trevisani for helpful discussions, Ryo Suzuki and Dmytro Volin for useful comments. The author is supported by the FCT fellowship SFRH/BPD/69813/2010. Centro de F\'isica do Porto is partially funded by FCT through the projects PTDC/FIS/099293/2008 and CERN/FP/116358/2010.

\end{document}